\documentclass{vgtc}                          

\graphicspath{{figures/}{pictures/}{images/}{./}} 

\usepackage{times}                     

\usepackage{tabu}                      
\usepackage{booktabs}                  
\usepackage{lipsum}                    
\usepackage{mwe}                       

\usepackage{mathptmx}                  
\usepackage{balance}

\onlineid{7999}

\vgtccategory{Research}

\vgtcinsertpkg

\title{Sensing Instability, Adapting the Scene: A Real-Time Movement-Smoothing Design Framework for Stable VR Locomotion}

\author{Ramisa Fariha Joyee\thanks{e-mail: ramisafariha.joyee123@gmail.com}\\ %
        \scriptsize Kennesaw State University %
\and M. Rasel Mahmud\thanks{e-mail: m.raselmahmud1@gmail.com}\\ %
     \scriptsize Kennesaw State University }

\teaser{
  \centering
  \includegraphics[width=\linewidth]{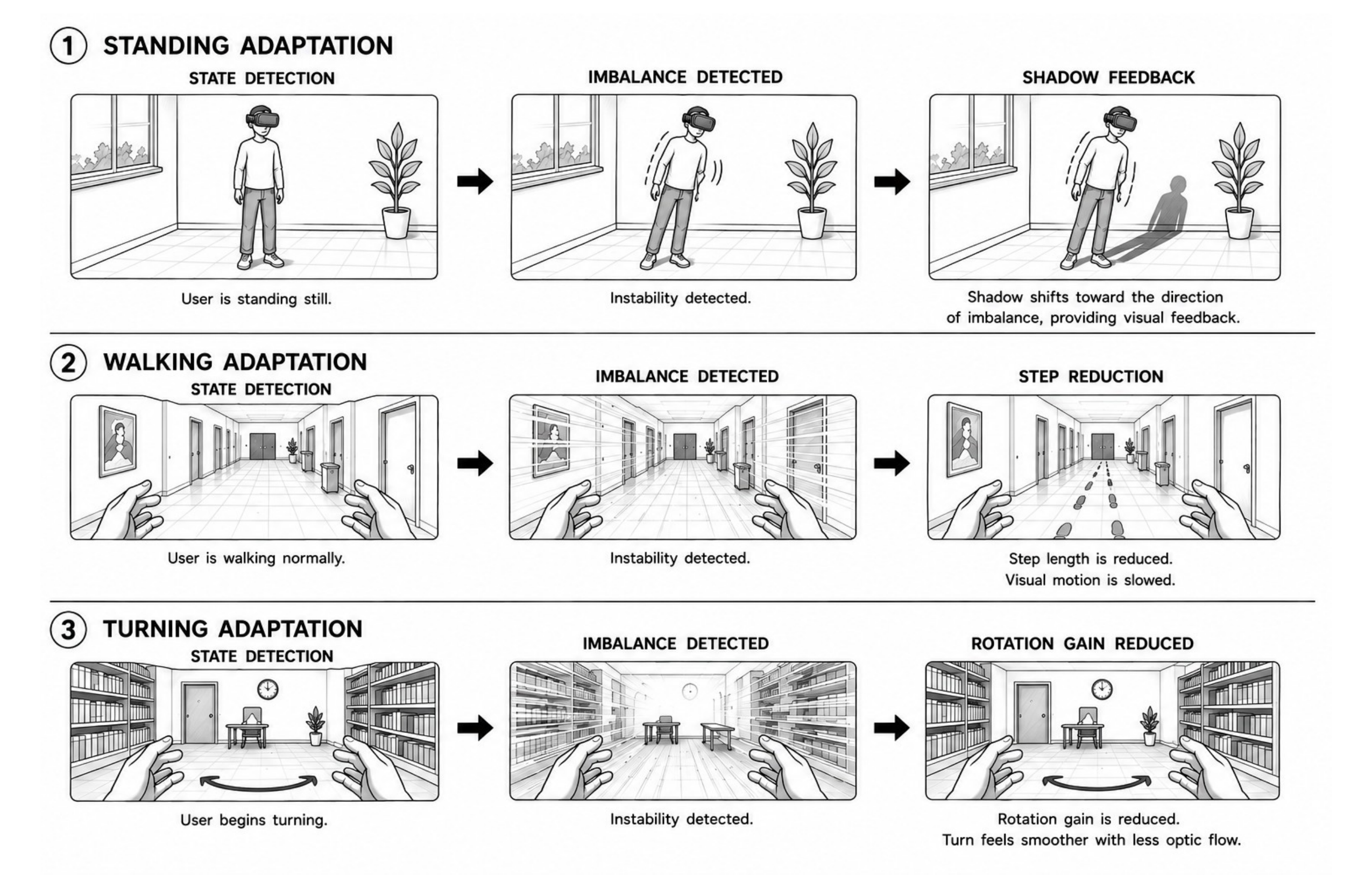}
  \caption{Overview of the proposed movement-smoothing adaptation framework. During standing, walking, and turning, the system continuously detects user instability and dynamically applies a movement-state-specific visual adaptation. (1) Standing is supported with a directional sway-responsive ground shadow, (2) walking with virtual step reduction, and (3) turning with yaw rotation gain reduction to improve postural stability during VR locomotion.}
  \label{fig:framework}
}

\abstract{
    Users experience different balance challenges while standing, walking, and turning in virtual reality (VR), yet most locomotion techniques apply the same visual behavior regardless of movement state. We present a real-time movement-smoothing design framework in this paper that organizes visual adaptations according to movement-specific balance demands. The framework introduces three design strategies targeting standing, walking, and turning, illustrating how state-aware visual adaptations can support postural stability during locomotion. This paper focuses on the framework design and implementation, while a user study is planned as future work. Our work guides the development of future context-aware VR locomotion systems that better support safe and comfortable navigation.
} 


\begin{document}



\maketitle

\section{Introduction}
    Virtual reality (VR) has become an important platform for applications such as gaming, education, training, healthcare, and rehabilitation, enabling users to interact naturally with immersive virtual environments \cite{calhau_2024_rehab, li_2020_training, oubibi_2022_teaching}. As these applications require users to navigate, explore, and perform tasks while moving, maintaining postural stability has become essential for ensuring both user safety and comfort. However, VR locomotion often introduces sensory conflicts because visual motion presented through the Head-Mounted Display (HMD) does not match the vestibular and proprioceptive information generated by the user's physical body. This mismatch can disrupt normal balance control, increase postural sway, and contribute to discomfort, disorientation, and even falls during immersive experiences \cite{mahmud_2022_standing, takahashi_2001_changebalance}. Consequently, improving balance during VR locomotion has become an important research problem.

    Previous studies have shown that visual information plays a critical role in maintaining human balance \cite{chander_2019_visualperturbations, duh_2001_fov}. Researchers have explored various sensory feedback techniques to improve user comfort and stability \cite{mahmud_2023_thesis, mahmud_2026_multimodal, mahmud_2023_investigating}, including real-time visual feedback for balance training \cite{mahmud_2023_theeyes, mahmud_2023_visualfeedback, mahmud_2024_multimdalfeedback, wang_2021_feedback}, auditory and vibrotactile feedback for postural control \cite{cordova_2023_realtime, hossain_2026_cogniaudit, mahmud_2022_auditory, mahmud_2022_standing, mahmud_2022_vibrotactile, mahmud_2025_vibrotactile}, locomotion gain adaptation \cite{teng_2025_effectofgain}, locomotion-aware foveated rendering \cite{shi_2023_foveated}, and controller movement strategies that reduce motion sickness during navigation \cite{tang_2025_strategies}. However, most existing approaches focus on applying a single adaptation throughout locomotion or optimizing a specific aspect of navigation, such as rendering efficiency, motion sickness, or interaction performance. In practice, locomotion is not a single continuous activity. Users repeatedly transition between different movement states, and each state places different demands on the balance system. During standing, maintaining stability primarily depends on continuously correcting natural body sway. During walking, users experience sustained forward optic flow that can increase visual-vestibular conflict. During turning, rapid rotational optic flow can disrupt spatial orientation and challenge postural control more severely than forward motion \cite{chander_2019_visualperturbations, peterka_2018_sensory}. Although these movement states differ in the mechanisms that contribute to instability, current VR locomotion techniques rarely distinguish between them, instead applying the same visual behavior regardless of how users are moving.
    
    We propose that movement-smoothing should be viewed as a movement-state-specific design problem rather than a single locomotion technique. Instead of relying on one universal adaptation, visual support should be designed according to the primary source of instability associated with each movement state. Based on this perspective, we present a movement-smoothing design framework that organizes visual adaptations according to movement-specific balance demands. The framework introduces three design strategies: restoring balance feedback during standing, reducing excessive forward optic flow and restoring balance during walking, and moderating rotational optic flow and restoring balance during turning. These strategies provide a conceptual foundation for designing future context-aware VR locomotion systems. Our goal is to provide a design framework that can guide the development of adaptive visual techniques for improving postural stability while preserving natural and immersive navigation experiences. In this paper, we focus on the framework design and implementation. A controlled user study to evaluate its effectiveness is left for future work.

\section{Design Framework}
    Our proposed framework considers locomotion as a sequence of distinct movement states, each associated with a different source of balance disturbance. We identify three common movement states encountered during controller-based VR locomotion, including standing, walking, and turning, and derive a corresponding design strategy for each state as shown in Figure \ref{fig:framework}. The framework consists of three components. First, the user's current movement state is continuously identified using locomotion information such as movement speed and rotational velocity via headset and controller motion. Instead of applying the same visual behavior throughout navigation, the framework classifies the user's activity into one of the three movement states. Walking is detected when translational velocity exceeds 0.05 m/s, and turning is detected when yaw velocity exceeds a threshold of 20 deg/s. These threshold values are set through pilot tuning to provide robust separation between standing, walking, and turning while minimizing false activations. To avoid rapid switching near the movement thresholds, the detected state remains active for at least 1 second before an adaptation is triggered. Once activated, the adaptation remains active until the corresponding movement signal falls below its threshold and the user returns to a stable state. Locomotion and rotational gains are then restored gradually, reducing abrupt visual changes during transitions between movement states. In the planned user evaluation, each participant will complete a short quiet-standing calibration to establish an individual baseline center of pressure (COP) via a BTrackS balance board (sampling rate 25 Hz) before the navigation task. A participant-specific deadzone is then determined through pilot tuning to distinguish normal postural sway from meaningful instability. When COP displacement exceeds this threshold, instability is detected, and the corresponding movement-state-specific adaptation is activated. During standing, maintaining postural equilibrium relies on continuous correction of natural body sway measured by a balance board. During walking, continuous forward motion generates optic flow that may increase visual-vestibular conflict and postural instability. Recent work has shown that providing supportive feedback during walking can improve gait performance and user stability in VR, motivating movement-specific adaptations during locomotion \cite{mahmud_2026_multimodal}. During turning, rapid rotational optic flow can disrupt spatial orientation and challenge balance more severely than straight movement. Finally, the framework selects a movement-specific visual adaptation designed to address the instability source while preserving a natural navigation experience.
    
    Unlike traditional locomotion techniques that employ a single adaptation regardless of user movement, the proposed framework emphasizes movement-state awareness. The objective is not to reduce all visual motion but to selectively support the balance system by addressing the particular challenge associated with the current movement state. This perspective allows different adaptation strategies to coexist within a unified design framework while remaining independent of their specific implementation. Based on this framework, we introduce three movement-smoothing design strategies, which are given below. 

    \subsubsection*{Standing state: Ground Shadow}
        When instability is detected during the standing state, a ground shadow appears and shifts proportionally toward the direction of body sway. The shadow provides continuous directional visual feedback of weight transfer. It may enable the user to make postural corrections while maintaining natural standing behavior.
    \subsubsection*{Walking state: Virtual Step Reduction}
        When instability is detected from COP displacement during walking, virtual locomotion speed is temporarily reduced. The speed is reduced by 50\%, which is determined by iterative pilot tuning. Smaller reductions had little noticeable effect on reducing visual motion, whereas larger reductions made locomotion feel less natural. The selected value provided a practical balance between reducing forward optic flow and preserving a natural navigation experience. This reduction lowers visual-motion velocity while preserving the user’s intended navigation path. Normal movement speed is restored gradually once stability improves.
    \subsubsection*{Turning state: Rotational Gain Reduction}
        During unstable turning, the rotational gain is temporarily reduced so that controller input produces slower scene rotation. The gain is reduced by 50\%, based on iterative pilot tuning. Smaller reductions produced little noticeable change in rotational optic flow, whereas larger reductions made turning feel sluggish. The selected value provided a practical balance between reducing visual motion and maintaining responsive, natural turning. The normal rotational gain is restored gradually once stability improves.
    
    Together, these strategies provide a conceptual foundation for designing future context-aware locomotion systems that adapt visual behavior according to movement-specific balance demands rather than relying on a single universal adaptation. We implemented the proposed framework in Unity version 6000.3.8f1 using the XR Interaction Toolkit and HTC Vive Focus Vision with a frame rate of 90 Hz. In the future user study, participants will navigate in a virtual environment containing the standing, walking, and turning adaptations as separate conditions. Although our work uses a balance board to estimate postural instability, the framework is independent of a specific sensing modality. Recent work has shown that instability and fall risk can also be estimated directly from integrated HMD and controller sensors, suggesting that the proposed framework is compatible with alternative sensing inputs \cite{pavel_2026_dtw, pavel_2025_vrfallnet}. Similar instability measures could be obtained from HMD motion, controller trajectories, or wearable inertial sensors, allowing the same movement-state-specific adaptation strategy to be applied across different VR hardware configurations.
    
\section{Conclusion and Future Work}
    We proposed a movement-smoothing design framework that introduces movement-state-specific visual adaptation strategies for standing, walking, and turning in VR. By matching visual interventions to movement-specific sources of instability, the framework offers a foundation for future context-aware locomotion systems. Future work will evaluate the framework's effect on postural stability, navigation performance, and user experience through a controlled user study. We will also investigate participant-specific instability thresholds and alternative sensing inputs derived from HMDs, controllers, and wearable devices. 

\balance
\bibliographystyle{abbrv-doi}

\bibliography{template}
\end{document}